**Speed of adaptation and genomic footprints of host-parasite coevolution under arms race and trench warfare dynamics**

Aurélien TELLIER[1]
Stefany MORENO-GÁMEZ[2]
Wolfgang STEPHAN[2]

[1] *Section of Population Genetics, Center of Life and Food Sciences Weihenstephan, Technische Universität München, 85354 Freising, Germany*

[2] *Section of Evolutionary Biology, LMU University of Munich, LMU BioCenter, Grosshaderner Str. 2, 82152 Planegg-Martinsried, Germany*

Corresponding author:
Aurélien TELLIER
*Section of Population Genetics, Center of Life and Food Sciences Weihenstephan, Technische Universität München, 85354 Freising, Germany*
Email: tellier@wzw.tum.de

**ABSTRACT**

Coevolution is expected to follow two alternative dynamics, often called "trench warfare" and "arms races" in plant-pathogen systems. "Trench warfare" situations are stable cycles of allele frequencies at the coevolving loci of both host and parasite, and it is predicted that the loci will show molecular evolutionary signatures of balancing selection, while "arms races" involve successive selective sweeps at the interacting loci. We study a haploid gene-for-gene model that includes mutation and genetic drift due to finite population size. We study the outcomes under different coevolutionary parameters to quantify the frequency of fixation of alleles, *i.e.* occurrence of an "arms race" dynamics. We find that contrary to the conventional wisdom, trench warfare situations do not imply larger numbers of coevolutionary cycles per unit time than arms races. Therefore, one cannot infer the nature of the dynamics in such systems based on the speed of coevolution estimated from cycle times. We subsequently perform coalescent simulations to generate sequences at the host and parasite loci. We ask whether the signatures expected under balancing selection or selective sweeps (unexpectedly high or low diversity, and high or low Tajima's D values, respectively) are likely to be observable in genomic data. Genomic footprints of recurrent selective sweeps are often found, whereas trench warfare yields signatures of balancing selection only in parasite sequences, and only in a limited parameter space with high effective population sizes ($N>1,000$) and long-term selection ($>4N$ generations). Therefore, the existence of a deterministic polymorphic equilibrium does not imply long-term trench warfare necessary for the signature of balancing selection to be observed in the coevolving genes' sequence. Our results suggest that to search for signatures of coevolution via population genomics, it is best to study pathogen rather than host genomes.

**Author summary**

Coevolution drives hosts to recognize pathogen molecules and mount an effective immune response, while pathogens evolve to avoid such recognition. We are interested in the evolutionary mechanisms driving such coevolutionary dynamics at host and pathogen genes of interaction. Two dynamics may occur which should exhibit different pace, and result in different genomic signatures in sequence data of host and pathogen genes: 1) the arms race in which host and pathogen populations accumulate advantageous alleles over short periods of time, and 2) the trench warfare in which several alleles cycle over long periods of time in both host and pathogen populations. In this study, we demonstrate that the classic expectations and genomic footprints of selection are only observed in a limited range of coevolutionary parameters. We also show in realistic theoretical models that the speed of coevolutionary cycles may often not differ between these dynamics, and is thus not a reliable measure to use in field studies. Finally, we demonstrate that it is more fruitful to sequence several pathogen genomes per population than host genomes in order to detect genes underlying coevolution. We urge for pathogen population genomic studies to discover new key genes for pathogen infectivity under coevolution.

# INTRODUCTION

Diseases are major agents of natural selection. In both natural and domesticated species, parasites limit plant and animal growth, alter development, and reduce seed and offspring production. There is selection pressure on hosts for resistance to parasites and equally on parasites to overcome host defenses. This confrontation drives coevolution, in which gene frequencies in one species determine the fitness of genotypes of the other species, and should result in genetic diversity for resistance and tolerance in hosts and in infectivity and virulence in parasites.

Progress in our molecular understanding of the genetic basis of resistance in hosts (humans, animals, or plants) and infectivity in pathogens (bacteria, fungi, viruses) reveal that few major defense genes underlie these traits [1,2,3,4]. Numerous theoretical analyses of gene-for-gene (GFG) or matching-allele models describe the coevolutionary dynamics at these loci based on the phenotypic outcome of infection determined by host genotype - parasite genotype (G×G) interactions. These models describe coevolutionary dynamics driven by negative indirect frequency-dependent selection (niFDS): rare alleles have a fitness advantage because selection in the host population depends on allele frequencies in the parasite population, and *vice versa* [5]. Two types of situations [6,7] are predicted to arise: 1) recurrent fixation of alleles and transient polymorphism, the so called "arms race" dynamics [8], or 2) continuous cycling of allele frequency changes, the "trench warfare" dynamics [9]. The trench warfare dynamics is also termed "Red Queen" dynamics in the animal-parasite literature [10]. Trench warfare has been predicted to exhibit faster coevolutionary cycles than the arms race [6,10]. It has been suggested that the occurrence of two dynamics can be determined empirically by measuring the reciprocal adaptation of the host and paathogens over time [11] in coevolving populations [12,13].

An alternative suggestion is that one can infer the type of coevolutionary dynamics by studying molecular evolutionary signatures in the sequences of host and parasite genes [6]. Based on deterministic models of coevolution with infinite host and parasite population sizes, the two types of dynamics should result in different observable signatures of polymorphism [6,7]. Arms races involve recurrent fixation of alleles in host and parasite populations, generating recurrent selective sweeps [14] at the coevolving genes [6,8]. The signatures of selective sweeps include an excess of low and high frequency variants (causing negative Tajima's D values in samples of sequences; [15]) and a region of reduced polymorphism around the selected site [14]. On the other hand, trench warfare situations [9] involve maintenance of polymorphism at the coevolving loci of both host and parasite, generating balancing selection [6,9]. Typically, this would be thought to produce high observed levels of polymorphism and an excess of intermediate frequency variants around the selected site, detectable by positive Tajima's D values [16].

With technological advances and the sequencing of numerous plant and animal genomes among and within populations, it has become feasible to detect genes under host-parasite coevolution [4,17,18,19]. However, searches for signatures of selective sweeps and balancing selection in host genomes have found few new genes whose sequences suggest that coevolution is occurring [17,18,19]. In fact, most genes underlying coevolution showing a genomic signature of selection had already been identified based on a limited set of candidate genes and sequenced individuals [8,9,20,21]. This lack of evidence for additional genes under coevolution may be due to the difficulty of disentangling the signatures of selection from demographic effects [22], but it is also worth examining the adequacy of the theoretical coevolutionary models, and asking whether they can predict observable patterns of molecular diversity. Note the importance to predict the polymorphism signatures at loci underlying coevolution, as scans for genomic footprints of selection have become useful methods to discover key pathogen genes for infection [23,24,25,26].

Links have intuitively been suggested between three levels of study of coevolution (represented in Figure 1): 1) the mathematical behavior of deterministic theoretical coevolutionary models, and specifically the stability of the polymorphic equilibrium (Figure 1, top graphs), 2) the occurrence of an arms race, when alleles are recurrently fixed in the populations (or the rarer alleles' frequencies are below a detection threshold), versus trench warfare, with alleles maintained in the host and parasite populations at observable frequencies (Figure 1, middle graphs), and 3) the incidence of selective sweeps or balancing selection (Figure 1, bottom graphs) inferred from population genomic data [6]. Under GFG models, trench warfare dynamics with long-term cycling of allele frequencies arises as random processes, genetic drift and mutation, in finite population size models nudge allele frequencies away from the stable polymorphic equilibrium point in the deterministic model ([27] and Figure 1). In matching-allele models, however, niFDS can generate stable limit cycles of allele frequency in host and pathogen populations.

Most models of coevolution have therefore assumed infinite population sizes and focused on the ecological and epidemiological conditions for the existence of a stable polymorphic equilibrium state [27,28,29]. In such models, costs of resistance and infectivity are necessary but not sufficient for stable polymorphic equilibria. A mathematical condition, negative direct frequency-dependent selection (ndFDS), is, however, required for stability of the polymorphic equilibrium point and is promoted by various life-history traits such as parasite polycyclicity [27,29]. These previous theoretical studies have not investigated the parameter conditions and model assumptions under which ndFDS generates trench warfare dynamics in a finite population, nor have they understood the links between stability of equilibria, coevolutionary dynamics and the occurrence of signatures of balancing selection expected at the interacting genes in natural populations.

To fill this major gap in the theory of host-pathogen coevolution, we study the behavior of biologically realistic models in populations of finite size and also generated polymorphism data at coevolving genes. We focus on the gene-for-gene (GFG) relationship which receives strong support from empirical studies in plants and their pathogens [1], as well as in invertebrates [2]. In the classical GFG model in a single population [28] the outcome of infection is determined by a single locus in haploid hosts and haploid parasites. At the host locus, two alleles, resistance (*RES*) and susceptibility (*res*) are present. The parasite has two alleles, for infectivity (*INF*) and non-infectivity (*ninf*). Infection occurs if the host is susceptible or if the parasite is infective. Note that in the plant pathology literature, the infectivity allele is called virulent and the non-infectivity is called avirulent [27]. We study GFG models without and with ndFDS [27]. First, "monocyclic" models with one parasite generation per host generation are universally unstable. In such models, all hosts are assumed to receive parasite spores at every generation, with frequency-dependent disease transmission. As niFDS generates only an unstable polymorphic equilibrium point [27], the monocyclic models are thus expected to always generate arms race dynamics in finite population (Figure 1). Second, the "polycyclic" models assuming several pathogen generations per host generation, are shown to generate ndFDS and thus a stable polymorphic equilibrium under some parameter values [27]. The "polycyclic" model yields thus the two possible states for the polymorphic equilibrium, stable or unstable, and can therefore generate the arms race or trench warfare dynamics depending on parameter values.

We compare finite population size versions of the monocyclic and polycyclic models. Based on intuitive expectations (Figure 1), the trench warfare dynamics should be commonly observed in the finite population size polycyclic model because this model exhibits a stable polymorphic equilibrium point over a wide range of parameter values. However, we find that stable polymorphism situations, with *sensu stricto* trench warfare dynamics, arise only with high costs of resistance and infectivity. The parameter space for which there is maintenance of polymorphism in the finite population polycyclic models is thus reduced compared to its infinite size version, due to the presence of genetic drift, which drives rare alleles to fixation. From our results, it also appears that for a wide range of parameter values the speed of the arms race dynamics, measured as the period of the cycles generated, is not slower than in trench warfare situations. We simulate in addition patterns of genetic polymorphism (SNPs) at host and parasite interacting loci using a coalescent simulator [30]. Footprints of recurrent selective sweeps are often found when occurring, but trench warfare outcomes yield signatures of balancing selection only in parasite sequences, and only in a limited parameter space, with high effective population sizes ($N$>1,000) and long-term selection (>$4N$ generations). We explain this in terms of coevolutionary cycles in which host alleles are prone to be present at low frequencies, and may therefore be lost from finite host populations. As a consequence, the existence of a deterministic polymorphic equilibrium does not imply the long-term trench warfare necessary for the signature of balancing selection to evolve in the coevolving genes' sequences. We show that footprints of coevolution at the genomic level are much more likely to be detected in pathogen than in host populations.

**RESULTS**

**Arms race versus trench warfare dynamics**

Four costs are associated with the simple GFG model. The costs of resistance ($u$) and infectivity ($b$) are fitness costs associated with the corresponding allele. Hosts can also exhibit a fitness cost of being diseased (denoted by $s$). Finally, $c$ is the cost to the non-infective parasite of not being able to infect the resistant host. We introduce for both models finite population sizes ($N$), mutation between *RES* and *res*, and between *INF* and *ninf* alleles ($\mu_{GFG}$), as well as the population mutation rate $\theta_{GFG} = 2N\mu_{GFG}$. Genetic drift is caused by binomial sampling based on allele frequencies at each generation. The polycyclic model is simple with two parasite generations per host generations [27]. Parasites undergo thus once genetic drift in the monocyclic model, but twice in the polycyclic model (equations in SI Appendix sections 1 and 2).

In the monocyclic model, which has an unstable polymorphic equilibrium point under all conditions, the main factor determining the speed of coevolution is the population mutation rate ($\theta_{GFG}$), while the costs ($u$, $b$, $s$) play a minor role. The speed increases with the mutation rate, and also depends on the population size (Figure S1) through two effects. $N$ determines the importance of genetic drift and thus the time to fixation of alleles and the times spent near fixation (Figure S2), while the population mutation rate ($\theta_{GFG}$) determines the rate at which new functional resistance and/or infectivity alleles arise in the population, and thus the rate at which coevolutionary cycles are initiated. Importantly, under the monocyclic model, fixation of resistance alleles can occur only in small populations and with intermediate to high costs of disease, *i.e.* only when selection for resistance, and drift, are both strong ($N < 5,000$, Figure S2). In the absence of costs ($u = b = 0$), infective and susceptible alleles become fixed. As expected from the deterministic monocyclic model [27], increasing values of the costs $u$ and $s$ move the polymorphic equilibrium away from the boundaries (SI Appendix section 1), corresponding to increased strength of selection for resistance and infective alleles. The speed of coevolution is also increased (Figures S1 and S2).

For the polycyclic model, genetic drift would be expected to be a key factor determining the transition between the trench warfare and arms race dynamics, in addition to the known effects of the cost values on the stability of the interior equilibrium point [27,28]. Our simulations with the same population mutation rate $\theta_{GFG} = 0.02$, but with different population sizes ($N$) demonstrate this (Figure 2). In this polycyclic model, smaller population sizes cause the allele frequencies to move further from the polymorphic interior equilibrium point, increasing the fixation probabilities even under high mutation rates. Stable long-term polymorphisms, defined as a near-zero probability of allele fixation, arises only for the larger population size modeled ($N = 10,000$), and only within a limited range of cost values (intermediate to high $u$ and $b$, low to intermediate $s$, Figure 2). Natural selection driving allele frequencies towards the stable polymorphic equilibrium point outweighs genetic drift. Trench warfare dynamics *sensu stricto* represents only a small proportion of the parameter space for which a stable interior polymorphic equilibrium occurs in the deterministic version of this polycyclic model (Figure 2).

We next compare the speed of coevolution between the two models, again assuming $\theta_{GFG}$ =0.02. With large populations ($N$ = 10,000) the arms race dynamics appears slower than trench warfare only when the costs $u = b$ are high (both around 0.2) and the cost of disease is intermediate ($s = \varphi = 0.5$; Figure 2). The difference in speed occurs for cost values under which fixation of alleles never occurs in the polycyclic model (so-called trench warfare *sensu stricto*, compare Figures 2 and 3). For small population size ($N$ = 1,000) evolution is only slightly faster in the polycyclic model for costs $u = b$ higher than 0.2 and $s$ higher than 0.3 (Figure S3). Our results narrow down the generality of the claim that arms races are always slower than trench warfare dynamics. In fact, for most parameter combinations, with the exceptions just mentioned, the speed of coevolution depends mainly on the population mutation rate ($\theta_{GFG}$).

**Genomic signatures of coevolution**

We generate expected genomic signatures at the host and parasite coevolution loci based on the allele frequency dynamics computed under the monocyclic and polycyclic models. We use the coalescent simulator *msms* [30] to generate neutral polymorphism segregating at these loci (parameters and set-up described below and in SI section 3 and Table S1). We summarize genomic effects of selection during coevolution in our models by two signatures, the genetic diversity measured as the number of segregating sites ($S$) and the Tajima's D ($D_T$), using the same population mutation parameter ($\theta_{GFG}$ = 0.02) and haploid population sizes as used above $N$ = 1,000 (Figures S5 and S6) and $N$ = 10,000 (Figures 6 and S7), with combinations of the other parameters, $u = b$ and $s = \varphi$. Examples of the distributions of Tajima's D obtained for two sets of parameters are shown in Figures 4 and 5 for four combinations of $N$ and $\mu_{GFG}$ values. Figures 6 and S5-S7 represent the average for all combinations of parameter values ($u = b$ and $s = \varphi$) of Tajima's D distributions seen in Figures 4 and 5.

As explained above, in small populations, trench warfare and thus balancing selection does not occur in the polycyclic model due to the action of genetic drift (Figures 4A-B, 5A-B and Figures S5 and S6). In the monocyclic model, both host and parasite populations have negative Tajima's D ($D_T <$ -1) under the range of cost values that yield stable long-term polymorphism with the larger population size modeled ($u = b$ are intermediate to high, and the cost of disease ($s$) is intermediate; see dark blue regions in Figures S5, S6). These regions represent parameter values for which recurrent selective sweeps occur (example in Figure 4D). As might be expected intuitively, the polycyclic model exhibits smaller magnitudes of negative Tajima's D values than the monocyclic model. However, smaller and probably more realistic costs ($u, b < 0.05$) yield site frequency spectra that do not appear different from that expected under neutrality ($D_T \approx 0$, light blue in Figures S5, S6 and examples in Figures 4A-C), because fixations of alleles occur very rapidly, so that host and parasite populations are monomorphic, or nearly so, for most of the time (Figures 1 and S1). As a consequence, the neutral SNPs at the ceovolutionary loci do not follow a structured coalescent. In the small population, the observed site frequency spectrum at neutral SNPs in the coevolutionary loci cannot be distinguished from that of loci in a population with no selection at nearby sites, this is especially true in host populations (see Figures 4A-B and 5A).

We next study an increased population size $N$, keeping $\theta_{GFG}$ constant (Figures 6 and S7). For the monocyclic model, similar negative values of Tajima's D, *i.e.* signatures of selective sweeps, are observed as for $N = 1,000$ (Figures 4D, 5C-D, and Figures 6 and S7). For this model, genetic drift is therefore not the main factor determining the genomic footprint ($D_T$). A second result is that high Tajima's D ($D_T > 1$), suggesting balancing selection, is observed only under conditions similar to those for small $N$ (intermediate to high values of costs $u = b$ and an intermediate cost of disease $s$; see Figures 6 and S7). Importantly, this signature is only observed in samples from the parasite population, but not in host sequences (examples in Figures 5C-D).

For the polycyclic model (Figures 6 and S7), balancing selection is indicated by positive $D_T$ values for parameter combinations that yield trench warfare *sensu stricto* in Figures 2 and 3. When the costs are small ($u$ and $b$ below 5%) this model does not generate a trench warfare dynamics or footprints of balancing selection. In fact, as allele fixation occurs under such cost values, the structured coalescent with two alleles in host and parasite occurs only over short periods of time. Importantly, in the finite population version of the polycyclic model, balancing selection is generated only in parasite populations, and only in a limited area of the parameter space (Figure 6), even though trench warfare dynamics occurs in host and parasite populations when $N=10,000$ (Figure 2). Comparing Figure 6 and Figure S7, we note that higher Tajima's D values ($D_T > 1$) are observed under balancing selection if the mutation rate is smaller. Finally, even if balancing selection is not clearly indicated by the host site frequency spectrum (as summarized by $D_T$), the situation can potentially be distinguished from neutrality by a higher genetic diversity, *i.e.* number of segregating sites, than expected under neutrality (Figure S8). Note that the signatures of different coevolutionary outcomes occur on different time scales. Signatures of recurrent selective sweeps, *i.e.* negative Tajima's D and smaller number of segregating sites than expected under neutrality, are observable in host and parasite sequences even if selection is as recent as $N$ haploid host generations (Figure S9). In contrast, balancing selection is detectable only if selection has been acting for at least $4N$ generations (Figure S8).

We explain these outcomes in the host and parasite populations as follows. First, trench warfare dynamics occurs for intermediate to high cost values of $u$ and $b$, which prevents fixation and maintains both host and parasite alleles in the populations. However, the equilibrium frequency of the host resistance (*RES*) allele is small ($\approx b$), so the effective population size of the *RES* type is small and this allelic class experiences regular bottlenecks due to random changes around the equilibrium frequency. Coalescent lineages associated with the *RES* type of the host locus may therefore often be lost. Mutational input from the *res* class, which has a higher effective size, homogenizes the sequences of the two types of alleles. Second, high amplitude of allele frequency changes may occur in the finite population model for some parameter values, so that the dynamics resemble more that of a series of incomplete sweeps than a structured coalescent. Thus, even if alleles never go to fixation under the trench warfare *stricto sensu*, such incomplete sweeps would reduce the signature of balancing selection. As a result, despite the maintenance of both *RES* and *res* types in the population, balancing selection (summarized here as $D_T > 1$) is not found in the host population (see examples in Figures 5C-D). In the parasite, bottlenecks due to frequency changes are less extreme, because the equilibrium frequencies for both allelic types are higher than for *RES* alleles in the host. At least two coalescent lineages are therefore maintained for each type (*ninf* and *INF*) over long periods of time, generating typical footprints of balancing selection in the parasite population.

**DISCUSSION**

A large body of the theoretical literature on coevolution, *e.g.* for GFG interactions, has focused on studying the stability of the internal polymorphic equilibrium in deterministic infinite population size models [27,28]. Our simulations show that the stability of the internal equilibrium in deterministic models is a sufficient, but not a necessary, condition for a trench warfare coevolutionary dynamics outcome. Moreover, it is naive to draw a direct link between signatures in sequences at host and parasite loci and the type of dynamics acting [6,29]. We study here a simple but widely applicable genetic model of host-parasite coevolution for genes with major phenotypic effects on the outcome of infection. In animal hosts, these genes may be *i*) upstream or downstream components of the innate immunity system (MHC, Interferons, Toll-like receptors in mammals [4]), and targets of parasite effectors [4,19], or *ii*) genes involved in RNA silencing pathways for virus resistance [2,20]. In plants, these are found to be *i*) involved in basal defense and non-host resistance [31], *ii*) intra-cellular targets of parasite effector molecules (guardee; [32]), or *iii*) R-genes interacting directly or indirectly with parasite effectors [1,9]. Our study highlights the value of modeling in planning stages before data collection in host and parasite populations, though this is rarely done in modern genomic studies. Predicting the situations for which balancing selection and selective sweeps signatures can be detected and distinguished from neutral sequence variants is important, because it illuminates what can and cannot be inferred from genome wide data of host and pathogens. Preliminary theoretical analyses help to determine the sample sizes and sampling schemes, *i.e.* how many populations and how many individuals per population to sample, that are likely to have reasonable power to distinguish the different coevolutionary situations based on knowledge of population structure, mutation rates and population sizes.

We find that balancing selection can occur only in a limited range of parameter values, and, even if it is occurring, will be detectable mainly in data from the parasite population, provided that population sizes are large enough (here $N > 1,000$). Our conclusion is consistent with previous understanding that, to be detected at a locus, balancing selection must have been very strong, acting for a sufficient number of generations, and with a low enough recombination rate, for private SNPs to occur between alleles [16,33]. This so-called structured coalescent is also observed at self-incompatibility genes which represent well known examples of balancing selection generated by direct frequency-dependent selection [16,34]. These expectations about the structured-coalescent are also valid for predicting footprints of coevolution in matching-allele models, which are used often in animal-parasite systems [3]. Our theoretical results support the findings of old balancing selection signatures, for example trans-specific polymorphisms at immunity loci in primates [35], between copies of a duplicated gene in plants [32]. We thus expect genome scans to be a promising way to detect balancing selection based on the site frequency spectrum in parasite genomes (using Tajima's D or Fay and Wu's H). In hosts, however, population genomics studies based on site frequency spectra will fail to discover genes under the kind of coevolutionary balancing selection modeled, because it will often create no large deviation from neutrality (see Figures 4, 5 and 6). However, scanning for significantly high nucleotide diversity over full genomes and at known defense genes, for example using the number of segregating sites and the Hudson-Kreitman-Aguade (HKA) test, may succeed (see Figure S8).

Our results shed thus light on the relative paucity of known immunity genes inferred to be under balancing selection, and of selective sweeps in host model organisms [4,18,19,21]. Selective sweeps, however, can be observed at genes undergoing recent coevolution, and may thus be observable in genes involved in resistance to infection by crop pathogens [24,25,36], particularly because the evolutionary dynamics in agriculture is recent and most likely follows a model with an unstable polymorphic equilibrium model [29]. This is consistent with the results of scans for positive selection that have been used to detect genes with novel functions [23,24]. Based on this time scale difference for the detection of selection, population genomic studies are thus much more able to detect arms races than trench warfare dynamics, even if the latter occur commonly in natural populations.

The rate of allele fixation in arms races is high when the costs of resistance ($u$) and infectivity ($b$) are realistically small (< 5%; [37,38] but see [39]). Two different fixation scenarios are of interest. First, the fixation of the resistance allele occurs only in small populations in both the polycyclic and monocyclic models, implying that natural plant populations should mostly be composed of susceptible individuals (as observed in [13]). We predict therefore that fixation and proliferation of resistance genes in the genome after duplication should occur in small populations under strong parasite pressure (high disease incidence and prevalence, and high cost of disease). Second, in parasite populations, fixation of infectivity occurs for a wide range of parameter values, favoring the increase in the number of effectors in parasite genomes by gene duplication [40].

By varying the population mutation rate ($\theta_{GFG}$) we disentangle in our study the influence of genetic drift and mutation on the maintenance of polymorphism in coevolutionary GFG models [41,42]. Mutations between types ($\mu_{GFG}$) prevent allelic fixation and shorten the waiting time for new alleles to arise in the population. Drift, on the other hand, has a different effect in the two models studied. In the monocyclic unstable model, it influences only the time that an allele spends at fixation (or near fixation), while in the polycyclic model it determines the probability of fixation. As a result, in this kind of model with a stable polymorphic equilibrium point, the maintenance of polymorphism when genetic drift occurs as well as mutation [42] is a necessary, but not sufficient condition, to generate observable footprints of balancing selection at coevolving genes, because the equilibrium allele frequencies are often small. Rather than a long-term polymorphism becoming established, all but one coalescent lineage is therefore often lost. Because the costs of the resistance (*u*) and infectivity (*b*) determine the allele frequencies at the deterministic equilibrium [28,29,38], they are key to generate observable non-neutral signatures in these models.

Contrary to previous suggestions [6,10], arms race dynamics in our GFG model may not be slower than under trench warfare. Defining the trench warfare dynamics *sensu stricto* by the absence of allele fixation in a finite population, fast coevolutionary cycles occur in the polycyclic model only for large population sizes, when the above costs are intermediate to high, and the cost of disease is intermediate (Figures 2 and 3). Attempts to distinguish between the two forms of coevolutionary dynamics using estimates of host and parasite population fitness at different time points [11,12] should therefore probably be restricted to this subset of the parameter values and to large populations. As coevolution in the arms race model becomes faster when increasing the mutational input ($\theta_{GFG}$), predicting the speed of the coevolutionary dynamics requires measuring both mutation rates and effective population size in host and parasite populations. We assume a one-locus GFG model but it may be more realistic to consider that several loci govern the interactions between hosts and parasites [42,43]). The deterministic versions of the used monocyclic and polycyclic models have been extended to multi-locus systems [43]. Under infinite population sizes, both multi-locus models behave similarly as their one-locus version regarding the stability of the polymorphic equilibrium point. However, the introduction of mutation and finite population size generates higher stochasticity in multi-locus models than in their one-locus counter-part, due to the higher number of alleles present in the population [42]. We suggest that footprints of balancing selection are not more likely under multi-locus models than shown here, however, as the per genome mutation rate will be higher with more loci, the speed of coevolution and number of coevolutionary cycles could be increased. This reinforces our conclusions that the speed cannot reliably be used to determine which coevolutionary dynamics occurs in a system of interest.

Under our conservative model assumptions, the classical expected genomic footprints of coevolution, selective sweeps or balancing selection, may often not be observed at coevolving loci. Combining several ecological and epidemiological characteristics which individually promote ndFDS may increase stability in a deterministic model [29] and potentially in finite population situations as well. If so, the likelihood of observing balancing selection may be higher in such situations. For example, we assume two parasite generations per host generation in our polycyclic model. Micro-parasites of invertebrates and fungal, viral and bacterial pathogens of plants undergo often much more generations per host generations. Increasing the polycyclicity of pathogens generates a higher likelihood of polymorphic equilibrium stability in deterministic models [27], but it has not been investigated if the corresponding increase in genetic drift in the pathogen population enhances the probability of allele fixation. We also assume here haploid hosts and parasites, whereas the diploid case is a more favorable situation for maintaining variability [44]. If many pathogen species are haploid, it is not a realistic condition for most plants. The GFG model of Leonard [28] can be stabilized when introducing diploid hosts [45] though the allele frequencies at the internal polymorphic equilibrium point are very similar to the haploid model. We thus anticipate that the effect of finite population size in models with diploid hosts will be very similar to the results shown here for both polycyclic and monocyclic models. Finally, our models assume for simplicity constant host and parasite population sizes, even though these sizes may vary in time due to random demographic events and density-dependent disease transmission. Such demographic changes are known random processes affecting the genome wide diversity and frequency spectrum, which would decrease the likelihood to detect genes under selection [22] in natural populations or experimental coevolution studies.

**METHODS**

**GFG models with finite population sizes**

Four costs are associated with the simple GFG model. The costs of resistance ($u > 0$) and infectivity ($b < 1$) are fitness costs associated with the corresponding alleles. Hosts can also exhibit a fitness cost of being diseased (denoted by $s$, with $0 < s < 1$). Finally, $c$ is the cost to non-infective parasites of being unable to infect the host (where $c \approx 1$ in GFG models). Recursion equations for the allelic frequencies and values of the equilibrium points for the monocyclic model can be formulated based on these fitness costs (SI Appendix Section 1, [27]). If the parasites undergo two generations per host generation (polycyclic model), the same parameters will be involved, but there can now be a stable equilibrium under some parameter conditions. As a plant grows between the two parasite generations, each new leaf may be infected by a spore produced either on the same plant or on another plant; these are called auto-infection and allo-infection, respectively [46]. We denote the auto-infection rate by $\psi$. The recursion equations and values of the equilibrium points are in the Section 2 of the SI Appendix [27]. The fitness costs of being diseased after one or two parasite generations are denoted by $\varepsilon$ and $\varphi$, respectively. For simplicity when comparing this model with the monocyclic one, we assume that $\varphi = s$, and that non-infective parasites cannot infect resistant hosts ($c = 1$). The stability of the polymorphic equilibrium point is determined by the strength of negative direct frequency-dependent selection (ndFDS; [27]; SI Appendix section 2). In the deterministic, infinite population version of this model, high auto-infection ($\psi = 0.95$) yields a stable equilibrium point over a wide range of the other parameter values. With $0.02 < s < 0.31$, this outcome arises when $u = b = 0.02$, and with $0.05 < s < 0.55$ it arises when $u = b = 0.05$ (see Figure 3 in [27]). Based on intuitive expectations (Figure 1), the trench warfare dynamics should be observed in the polycyclic model over a wide range of parameter values.

To study these two models in finite populations, we assume host and parasite populations with sizes $N_H$ and $N_P$, respectively, and genetic drift is introduced by binomial sampling based on allele frequencies in each host or parasite generation. The population sizes are assumed to be constant over time, and for simplicity in our simulations $N = N_H = N_P$. Mutations between allelic types are introduced from *RES* to *res* or *INF* to *ninf* alleles and *vice versa* [41], with forward and backward mutations having the same rate $\mu_{GFG}$ per generation and their number following a Poisson distribution. Mutation ($\mu_{GFG}$) and population size ($N$) determine the population mutation rate parameter $\theta_{GFG} = 2N\mu_{GFG}$ which defines the rate of appearance of new alleles (assumed to be the same in both host and pathogen populations). To disentangle the influence of mutation on the dynamics of allele frequencies from that of drift, we compare characteristics of the coevolutionary dynamics which are relevant for population genetics (see below) using fixed values of the population mutation parameter ($\theta_{GFG}$), with varying population sizes from $N = 1,000$ (which we call small) to $N = 10,000$ (large), and mutation rate ($\mu_{GFG}$ values of $10^{-4}$, $10^{-5}$ and $10^{-6}$). Table S1 summarizes all the parameters. We also relax the assumption of equal host and parasite population sizes, as this may be more realistic, and show that the results are not quantitatively affected (Figure S4).

**Statistical analysis of the dynamics**

We simulate the allele frequency dynamics under both monocyclic and polycyclic models over 10,000 host generations using R scripts. We measure the percentage of time that host or parasite alleles are fixed. As high mutation rates prevent the fixation of alleles, even if the polymorphic equilibrium is unstable [41,42], we compute the percentage of time that allele frequencies are within 5% of fixation or extinction. The speed of coevolution is measured by counting the total number of cycles for both the resistance and infectivity alleles over 10,000 generations. A cycle is defined in the trench warfare as the period in generations between two consecutive maximum values of the host (or parasite) allele frequencies. For arms races, it is the time between successive allele fixations. These values are computed by fitting a smooth spline curve to the allele frequency trajectories with the "smooth.spline" function from the R package "stats" (with smoothing parameter equal to 0.15, after testing various smoothing parameters from 0.05 to 0.5 to find the value which ensures the greatest robustness and accuracy). The allele frequencies at the start of simulations may affect the behavior of the system, due to the existence of unstable limit cycles [41]. Although we do not expect to find limit cycles in our models, all statistics are averages over 100 runs with varying initial frequencies sampled from a uniform distribution ($a_0$ and $R_0$ in the interval [0.01, 0.5]). We study the coevolutionary dynamics over a biologically plausible range of the cost parameters (assuming $u = b$) by varying between no costs (0) to high costs (0.3), and allowing costs of disease $s$ and $\varphi$ to range from low (0.01) to high (0.6) values (Table S1).

**Generating expected genomic signatures**

We assume a constant population size, $N$, and also that the *INF* or *RES* types are each caused by a single SNP located in the center of each locus. We use the coalescent simulator *msms* [30] to generate neutral polymorphisms at other sites at these loci. We first use the R scripts described above to generate the host and parasite allele frequency dynamics for a given parameter combination, assuming that selection acts over $6N$ haploid host generations. This is referred to as the path of allele frequency. Allele frequency paths are generated assuming initial allele frequencies of $R_0 = a_0 = 0.1$. Individual paths are used as inputs for the *msms* program, to generate a neutral coalescent tree backward in time for each locus under the allele frequency path given and sample size $n$.

We obtain sequences for sample size of $n = 40$ haploid hosts and $n = 40$ haploid parasites. Similar results are obtained with larger sample size of $n = 200$ (SI Appendix Section 3). The haploid host and parasite population sizes are set to $N_P = N_H = N = 1,000$ and 10,000, and two mutation rates are defined for the coalescent simulations. First, the coevolution mutation rate $\mu_{GFG}$ defined above is the rate at the polymorphic site determining the coevolving types (*RES* and *res*, and *INF* and *ninf*), and we assume the same values of $10^{-4}$, $10^{-5}$ and $10^{-6}$ for both forward and back mutations. Second, we define a neutral mutation rate, $\mu_{neutral}$, for neutral SNPs within each locus. The neutral population mutation rate is set to $\theta_{neutral} = 2N\mu_{neutral} = 20$ per locus. For $N = 1,000$ and assuming a locus length of 2kb, the mutation rate is therefore $\mu_{neutral} = 2.5 \times 10^{-6}$ (for $N = 10,000$, $\mu_{neutral} = 2.5 \times 10^{-7}$). This mutation rate is chosen to generate approximately 85 segregating sites per locus for a sample of size $n = 40$ under a neutral model without selection, which allows us to be confident in the statistical comparisons between loci. Smaller neutral mutation rates would generate smaller numbers of SNPs and decrease the statistical power to distinguish between different genomic signatures. Using a set of C++ scripts, we compute summary statistics from our samples including the number of segregating sites ($S$), the site frequency spectrum and Tajima's D ($D_T$, [15]). The number of segregating sites is a measure of genetic diversity at a locus, and Tajima's D is a summary of the site frequency spectrum which is commonly used to detect loci under selection as positive $D_T$ values indicate possible selective sweeps and negative $D_T$ values indicate possible balancing selection.

For a given parameter combination, two types of stochasticity occur in our simulations: 1) stochasticity among allele frequency paths due to genetic drift and randomness of mutations, and 2) stochasticity of the coalescent process for a given frequency path. Preliminary analyses show that the variability in genomic signatures is greater among frequency paths than among replicates of the coalescent process. We therefore simulate 2,000 host and parasite frequency paths for each parameter combination, with one coalescent simulation per path. The mean of the distributions of $S$ and $D_T$ over the 2,000 simulations is computed to examine the potential footprints of polymorphism or fixation at the locus. An important point, which we make use of below, is that, like other cases with balancing selection (which can helpfully viewed as cases of population subdivision or structure), our coevolution model is a form of structured coalescent. In the case we study, there are two functional types in each host and parasite population (host *RES* and *res*, and parasite *INF* and *ninf*), and the types are linked by mutation between them at rate $\mu_{GFG}$ (a form of migration between the different functional lineages). Within the host and parasite populations, the allele frequencies vary in time during the coevolutionary process, and this generates a varying population size for each functional allelic type.

Preliminary analyses are used to evaluate the simulation conditions (SI Appendix Section 3). The major determinants of observable genomic signatures are the strength of selection, and the time during which selection occurs [33]. As our aim is to compare the genomic signatures for different strength of selection (coevolution), the time of selection is fixed for all simulations to $6N$ haploid host generations. Under a simple model of strong balancing selection with two allelic types at a fixed frequency of 0.5, an excess of intermediate frequency variants in the site frequency spectrum (and high $D_T$ values) are only observed when selection occurs for at least $4N$ generations (SI Appendix Section 3). This corresponds to the necessary time for the structured coalescent to generate distinct SNP frequencies within allelic types. An increase of migration between allelic types due to mutation ($\mu_{GFG}$) or intra-locus recombination ($\rho$) decreases the difference in neutral allele frequencies between types, and $D_T$ converges towards zero as expected under neutrality. After preliminary analyses (SI Appendix Section 3), to be conservative, we choose to simulate coevolution over $6N$ host generations, and assume no intra-locus recombination ($\rho = 0$).

## ACKNOWLEDGMENTS

AT acknowledges support from DFG grant HU 1776/1 to Stephan Hutter and the German Federal Ministry of Education and Research (BMBF) within the AgroClustEr "Synbreed – Synergistic plant and animal breeding" (FKZ: 0315528I). WS was funded by DFG grants HU 1776/1 and STE 325/14. SM was funded by the European Union through the Erasmus Mundus Master Program in Evolutionary Biology.

**FIGURE LEGENDS**

**Figure 1:** Connecting model expectations for the arms race and trench warfare coevolutionary dynamics. Expectations for the host resistance and parasite infectivity allele are in red and blue, respectively. On the left side, an unstable internal polymorphic equilibrium point in the deterministic model with infinite population size model shows allele frequency cycles with increasing amplitude. This yields a series of recurrent fixations of host and parasite alleles, the arms race dynamics, under finite population size with genetic drift and mutation. This dynamics is suggested to result in recurrent selective sweeps which are detected by negative values of Tajima's D at these coevolutionary loci in comparison to the neutral genomic background. The expected value at a given host or parasite coevolutionary locus is drawn from the red and blue distributions, while the expected neutral distribution of Tajima's D over the genome is centered on zero (dotted black line). On the right side, a stable internal polymorphic equilibrium point in the deterministic model shows allele frequency cycles with decreasing amplitude. This yields a series of stochastic frequency cycles of small amplitude around the equilibrium value, the trench warfare dynamics, under finite population size with genetic drift and mutation. This dynamics is suggested to result in long-term balancing selection which is detected by positive values of Tajima's D at these coevolutionary loci in comparison to the neutral genomic background.

**Figure 2:** Percentage of time that host susceptible alleles are fixed or near fixation in the polycyclic model 2. The percentage of fixation time (allele frequency > 95%) is a function of the costs of resistance and infectivity ($u = b$) and cost of being diseased ($s = \varphi$). Other parameters are: $c = 1$; $\psi = 0.95$. The GFG population mutation rate is fixed ($\theta_{GFG} = 0.02$), but population size $N$ varies ($N = 1,000$ in the left graph, $N = 10,000$ in the right graph).

**Figure 3:** Number of coevolutionary cycles under the monocyclic (left graph) and polycyclic (righ graph) models. The number of cycles is computed per 10,000 host generations as function of the costs of resistance and infectivity ($u = b$) and cost of being diseased ($s = \varphi$). Other parameters are: $c = 1$; $\psi = 0.95$. The GFG population mutation rate is fixed ($\theta_{GFG} = 0.02$), $N = 10,000$ and $\mu_{GFG} = 10^{-6}$.

**Figure 4:** Tajima's D distribution at host and parasite loci for both models. The host distribution is in blue, the parasite in red. Outcomes of the monocyclic model are represented as dotted lines and those of the polycyclic model as solid lines. Different population mutation rates are chosen with different population size ($N$) and mutation rate ($\mu_{GFG}$). Each distribution is based on 2,000 repetitions. The parameters are $u = b = 0.05$; $s = \varphi = 0.36$; $c = 1$; $\psi = 0.95$; $\theta_{neutral} = 20$; selection acts for $6N$ generations and the sample size is $n = 40$ haploid hosts and haploid parasites. A) $N=1,000$ and $\mu_{GFG} = 10^{-4}$; B) $N=1,000$ and $\mu_{GFG} = 10^{-5}$; C) $N=10,000$ and $\mu_{GFG} = 10^{-5}$; D) $N=10,000$ and $\mu_{GFG} = 10^{-6}$.

**Figure 5:** Tajima's D distribution at host and parasite loci for both models. The host distribution is in blue, the parasite in red. Outcomes of the monocyclic model are represented as dotted lines and those of the polycyclic model as solid lines. Different population mutation rates are chosen with different population size ($N$) and mutation rate ($\mu_{GFG}$). Each distribution is based on 2,000 repetitions. The parameters are $u = b = 0.2$; $s = \varphi = 0.36$; $c = 1$; $\psi = 0.95$; $\theta_{neutral} = 20$; selection acts for $6N$ generations and the sample size is $n = 40$ haploid hosts and haploid parasites. A) $N=1,000$ and $\mu_{GFG} = 10^{-4}$; B) $N=1,000$ and $\mu_{GFG} = 10^{-5}$; C) $N=10,000$ and $\mu_{GFG} = 10^{-5}$; D) $N=10,000$ and $\mu_{GFG} = 10^{-6}$.

**Figure 6:** Mean of Tajima's D ($D_T$) for the monocyclic (left graphs) and polycyclic (right graphs) as function of the costs of resistance and cost of infectivity ($u = b$) and the cost of being diseased ($s = \varphi$). Population size is $N = 10,000$, mutation rate is fixed to $\mu_{GFG}=10^{-5}$ (so $\theta_{GFG} = 0.2$), $c = 1$; $\psi = 0.95$; $\theta_{neutral} = 20$; selection acts for $6N$ generations and the sample size is $n = 40$ haploid hosts and haploid parasites. A) host locus, and B) parasite locus.

Figure 1

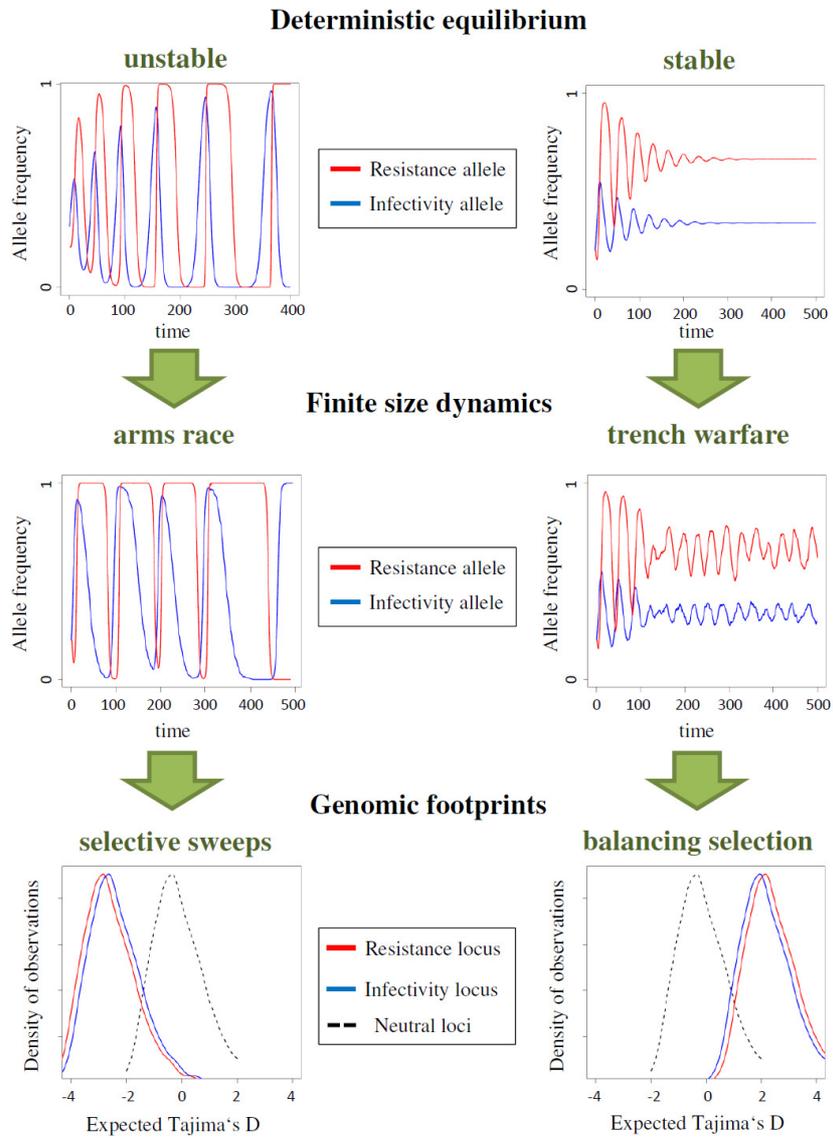

Figure 2

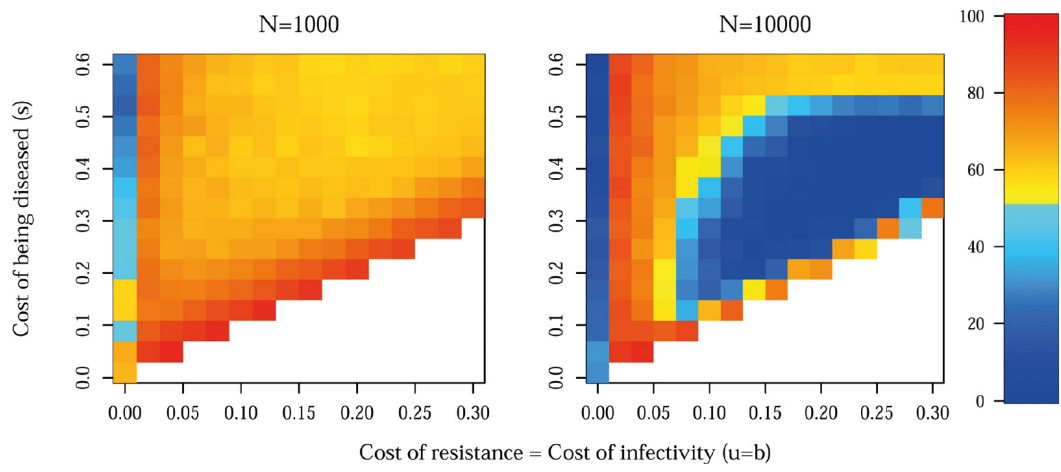

Figure 3

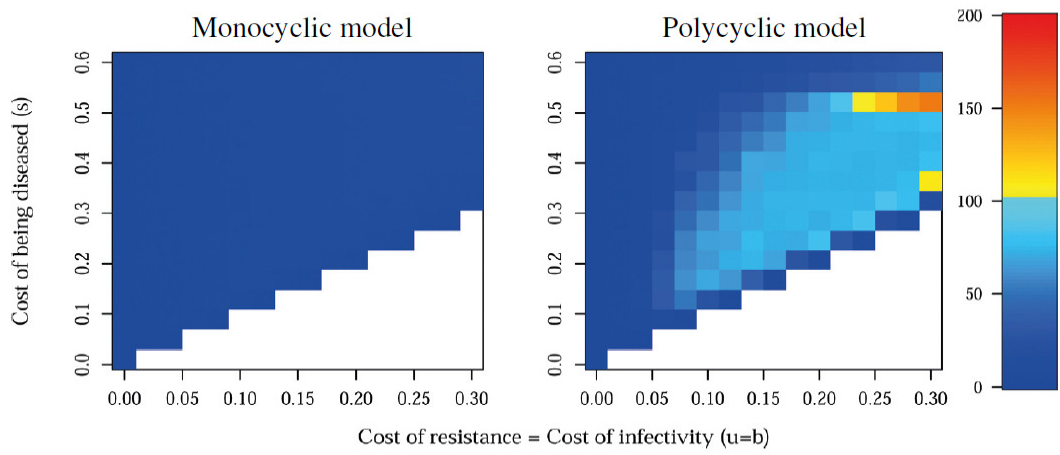

Figure 4

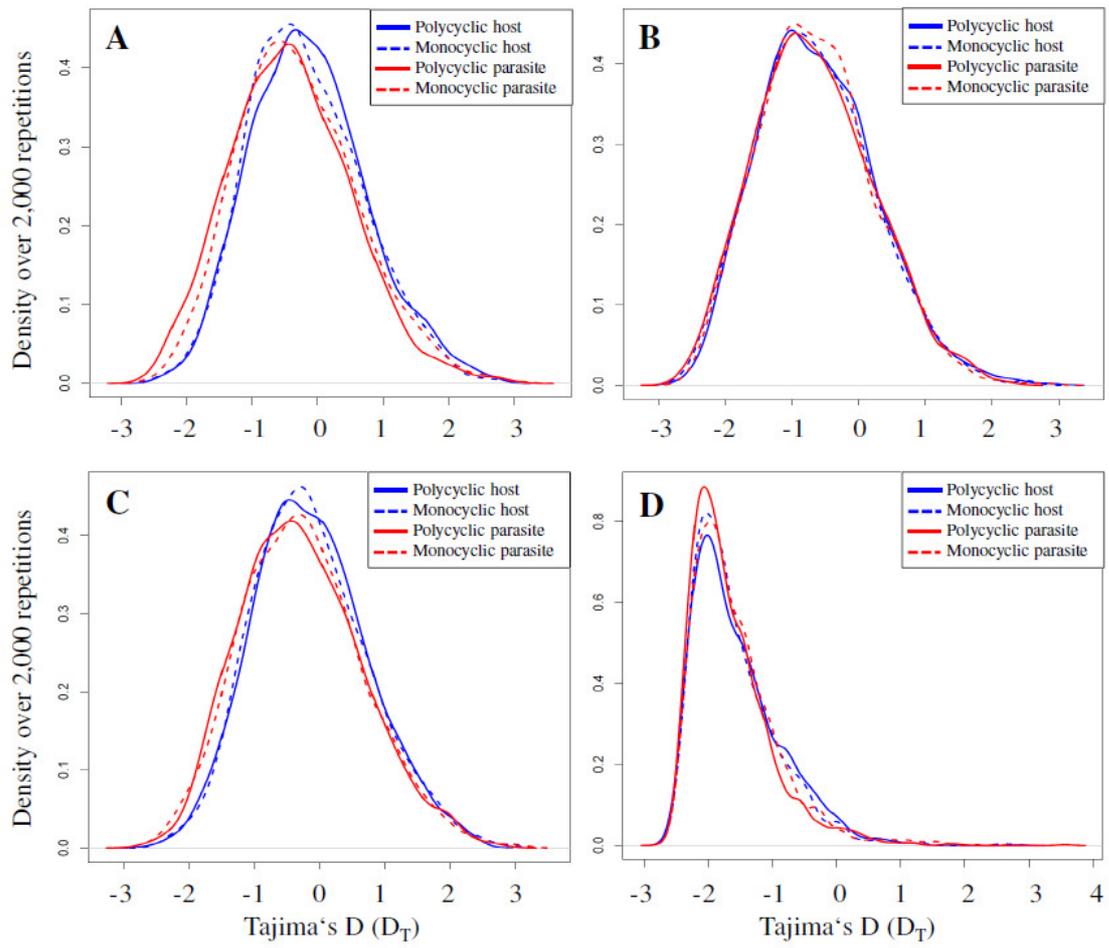

Figure 5

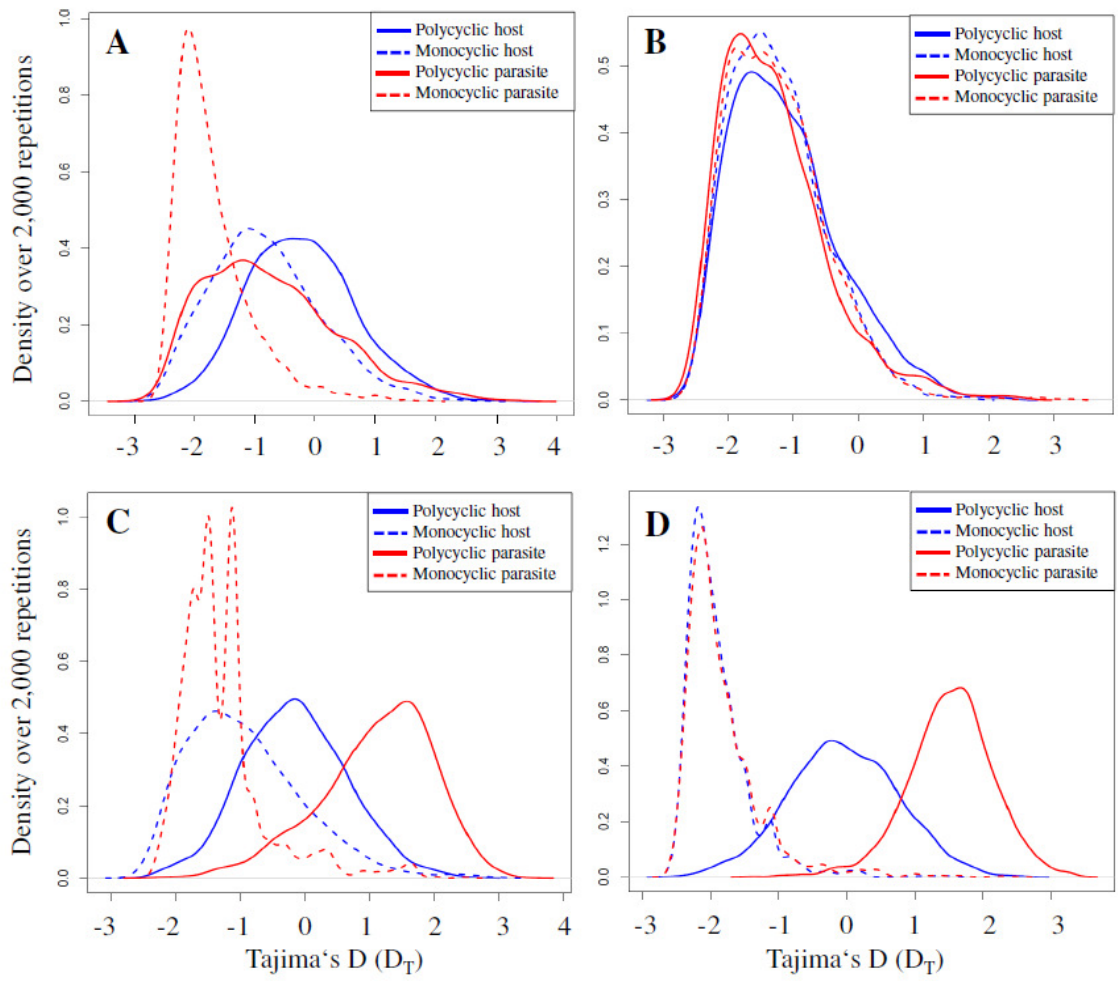

Figure 6

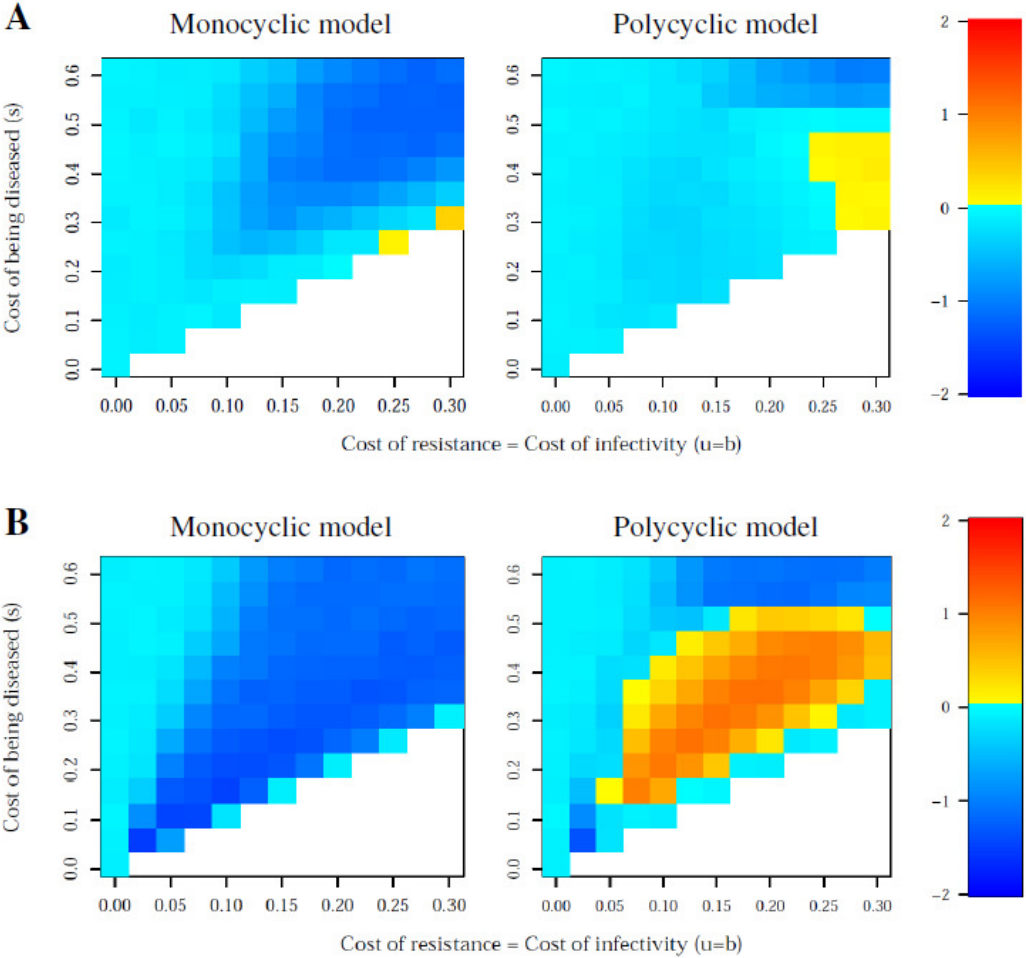